\documentclass[10pt,a4paper]{article}

\usepackage{indentfirst}			
 \usepackage{enumerate}				
\usepackage{amsmath, amsgen,amsfonts,amssymb,amsbsy}	

\begin{document}

\thispagestyle{plain}		

\title{Einstein Spacetimes with Constant Weyl Eigenvalues}
\author{Alan Barnes \\ 
School of Engineering \& Applied Sciences, \\
Aston University,\\
Birmingham B4 7ET, \\
United Kingdom.}

\maketitle

\begin{abstract}
Einstein spacetimes (that is vacuum spacetimes possibly with a non-zero cosmological constant $\Lambda$) with constant  non-zero Weyl eigenvalus are considered. 
For type Petrov II \& D this assumption allows one to prove that the non-repeated eigenvalue necessarily has the value $2\Lambda/3$ and it turns out that the only possible spacetimes are some Kundt-waves considered by Lewandowski which are type II and  a Robinson-Bertotti solution of type D.  

For Petrov type I  the only solution turns out to be a homogeneous pure vacuum solution found long ago by Petrov using group theoretic methods. These results can be summarised by the statement that the only vacuum spacetimes with constant Weyl eigenvalues are either homogeneous or are Kundt spacetimes. This result is similar to that of Coley et al.\ who proved their  result for {\bf general} spacetimes under the assumption
that all scalar invariants constructed from the curvature tensor {\bf and all its derivatives} were constant.

Some preliminary results are also presented for Petrov Type I vacua in which either only one of the Weyl eigenvalues is constant or in which the ratios of the Weyl eigenvalues are constants.  In particular in each case there is a simple algebraic relation between the  Newman-Penrose Weyl tensor components $\Psi_2$ \& $\Psi_0$ (=$\Psi_4$) and the `cross-ratio' of the optical scalars $\kappa\nu-\sigma\lambda$ of the associated principal null tetrad of the Weyl tensor.
\end{abstract}

\vspace{-5 pt}
\section{Introduction}
\noindent In this paper special classes of solutions of Einstein's vacuum field equations in general relativity, possibly with a non-zero cosmological constant 
$\Lambda$,
\begin{equation}
R_{ab} = \Lambda g_{ab}
\label{efe}
\end{equation}
are considered.  In the solutions considered it is assumed that the eigenvalues of the Weyl tensor, regarded as a a linear transformation acting on the six-dimensional  space of bivectors, satisfy certain restrictions.  The eigenvalues $\lambda$  and eigenbivectors $V^{ab}$ of the Weyl tensor satisfy the equation
\begin{equation}
C^{ab}_{\ \ \ \ cd}V^{cd} =2\lambda V^{ab}
\label{ev-eq1}
\end{equation}
where the factor two, which arises due to the contraction over a pair of antisymmetric indices, is included for later convenience.  Owing to the duality conditions satisfied by the Weyl tensor there are essentially only three roots of the characteristic equation and these sum to zero due to the trace-free nature of the Weyl tensor (see, for example, chapter 4 of Stephani et al.\cite{exact-sol} for more details).

In this paper three distinct classes of fields are considered: 
\newcounter{N}
\begin{list}
{(\roman{N})}{\usecounter{N}
       \setlength{\rightmargin}{\leftmargin}}
  \item Petrov types I, II \& D fields where the Weyl eigenvalues are all constant;
  \item Petrov type I fields where one of the Weyl eigenvalues is a constant;
  \item Petrov type I  fields where all the Weyl eigenvalues are constant multiples of a single (complex) scalar field.
\end{list}
The conformally flat fields are of constant curvature and need not be considered further as they are well understood. Petrov types III and N  fields are not considered in this paper; assumptions on the constancy or linear dependence of the Weyl eigenvalues do not restrict these fields as all the eigenvalues are necessarily zero. Similarly we do not consider  Petrov types II \& D fields of class (iii) as the assumption of linear dependence of the eigenvalues is necessarily satisfied.   For Petrov types II \& D the assumptions on the eigenvalues in class (i) and (ii) are, of course, equivalent.

The subclass of Petrov type I fields of class (i) can be regarded as the intersection of the classes (ii) \& (iii).   Furthermore for Petrov type I fields it may be assumed that the Weyl eigenvalues are non-zero since Brans\cite{brans} showed that pure vacuum Petrov type I fields with a zero Weyl eigenvalue do not exist. His proof generalises trivially to the case when $\Lambda$ is non-zero.

There are a number of motivations for investigating fields of this sort.  Firstly, most known vacuum solutions of the field equations are either algebraically special or admit a group of isometries and so it would be interesting to enlarge the class of known vacuum solutions of Petrov type I. Making simplifying assumptions regarding the scalar curvature invariants seems to be a promising approach which has hitherto been largely neglected. In particular the assumptions for classes (ii) \& (iii) can be seen as natural generalisations of Brans' assumption of a zero Weyl eigenvalue; the vanishing of an eigenvalue, of course, implies that the eigenvalues satisfy assumption (iii) due to the zero trace condition on the Weyl tensor in four dimensions.  Similarly the assumption (iii) can been seen as the extension to Petrov type I fields of a condition that holds automatically for fields of Petrov types II and D.

Secondly, in recent years there has been considerable interest in general spacetimes in which all the scalar invariants constructed from the curvature tensor and its covariant derivatives are constant.  For example Coley and his collaborators\cite{coley} showed that such spacetimes were either homogeneous or are a special class of Kundt-waves\cite{kundt}.  In this paper it is shown that a similar result holds for four dimensional Einstein spaces under the weaker assumption that the scalar curvature invariants constructed from the curvature tensor {\bf alone} are constant and not all zero.  It would be interesting to see how far this result could be extended to non-vacuum spacetimes.

Thirdly, if we consider Einstein spacetimes of embedding class two (that is spacetimes that can be locally isometrically embedded  in a flat six-dimensional pseudo-Riemannian manifold) the commutator  $C^{ab}=\Omega^{[a}_c \Lambda^{b]c}$ of the two second fundamental forms
 $\Omega^{ab}$ and $\Lambda^{ab}$ associated  with the embedding satisfies Yakupov's identity\cite{yak1}
\begin{equation}
C^{ab}_{\ \ \ \ cd}C^{cd} =4\Lambda/3 C^{ab}
\label{yakid}
\end{equation}
Thus, if the commutator is non-zero, such spacetimes necessarily have a constant Weyl eigenvalue $\lambda=2\Lambda/3$.  Remarkably it turns out that if we consider Einstein spacetimes of Petrov types II and D with constant Weyl eigenvalues, the value of the non-repeated eigenvalue is necessarily $2\Lambda/3$; no other values are permitted. Thus the class of these fields of Petrov type II \& D include all the algebraically special proper Einstein spaces of embedding class two with torsion.

In fact Yakupov\cite{yak1, yak2} claimed that the commutator $C^{ab}$ necessarily vanished and so the torsion vector of the embedding was a gradient and so could be set to zero by suitable choice of the two normal vectors to the spacetime in the enveloping six-dimensional space.  However, no proof of this claim was ever published.  This claim, if true, would greatly simplify the analysis of embedding class two Einstein spacetimes as the Ricci equation is identically satisfied and the two Codazzi equations decouple.  However, it is easy to find very many examples of pairs of $\Omega^{ab}$ and $\Lambda^{ab}$ which satisfy the Einstein space conditions, but which do not commute. In order to prove Yakupov's claim it would suffice to show that none of these pairs can satisfy the Codazzi and Ricci integrability equations. Conversely if any pair were found to satisfy these integrability conditions it would constitute a counterexample to Yakupov's claim.  This problem is currently under investigation by the author and will be reported upon elsewhere\cite{ab1}.

\section{Petrov Classification}
\noindent In this section some useful results on the Petrov classification which will be needed later in the paper, are briefly reviewed. The conditions on the Weyl eigenvalues are also expressed in the notation of the Newman-Penrose formalism\cite{NP} which will be used in the analysis later in the paper. These results are mostly adapted from chapters 3 and 4 of Stephani et al.\cite{exact-sol}.  
 
 By contracting the eigenvalue equation  \eqref{ev-eq1} a unit timelike vector $u^a$ it can be cast into the equivalent form
 \begin{equation}
Q_{ab}X^b =\lambda X_a
\label{ev-eq2}
\end{equation}
where $Q_{ab} = E_{ab}+iH_{ab} \equiv (C_{acbd}+i^{\ *}C_{acbd})u^cu^d$ and $E_{ab}$ and $H_{ab}$ are real symmetric trace-free tensors orthogonal to $u^a$ which are known as the electric and magnetic parts of the Weyl tensor respectively\cite{matte}.  Introducing an orthonormal tetrad $(e_1^a, e_2^a, e_3^a, u^a)$ and an associated complex null tetrad $(\ell^a, n^a, m^a, \bar m^a)$ defined by 
\begin{equation}
\ell^a=\tfrac{1}{\sqrt{2}}(u^a+e_3^a),\quad n^a=\tfrac{1}{\sqrt{2}}(u^a-e_3^a), \quad m^a=\tfrac{1}{\sqrt{2}}(e_1^a+ie_2^a)
\label{nul-tet}
\end{equation}
the orthonormal frame components of $Q_{ab}$ may be written in the form 
 \begin{equation}
{\bf Q}= \begin{pmatrix} \Psi_2-\tfrac{1}{2}(\Psi_0+\Psi_4) & \tfrac{1}{2}i(\Psi_4-\Psi_0) &\Psi_1-\Psi_3\\ 
\tfrac{1}{2}i(\Psi_4-\Psi_0) &\Psi_2+\tfrac{1}{2}(\Psi_0+\Psi_4)& i(\Psi_1+\Psi_3) \\
\Psi_1-\Psi_3 &i(\Psi_1+\Psi_3)& -2\Psi_2\end{pmatrix}  
\label{gen-Q}
\end{equation}
where the $\Psi_i (i =0\ldots 4)$ are the five complex Newman-Penrose components of the Weyl tensor \cite{NP}.

By a suitable choice of the orthonormal tetrad $(e_1^a, e_2^a, e_3^a, u^a)$, the complex matrix $\bf Q$ can be reduced to one of six canonical forms depending on the elementary divisors and multiplicities of the eigenvalues of $\bf Q$.  These correspond to the six Petrov types \cite{petrov1} namely I, II, D, III, N and O.  For Petrov types I and D there is an orthonormal tetrad  such that  ${\bf Q}$ is diagonalisable:
$${\bf Q}= \begin{pmatrix} \lambda_1 & 0 &0\\0 &\lambda_2 & 0 \\0 &0& \lambda_3\end{pmatrix} 
\qquad \lambda_1+\lambda_2+\lambda_3=0$$ 
where for type D: $\lambda_1=\lambda_2= -\lambda_3/2$.  Thus, in the associated null tetrad, the NP Weyl tensor components satisfy
\begin{equation}
\Psi_0=\Psi_4 =(\lambda_2-\lambda_1)/2 \qquad \Psi_1=\Psi_3 = 0\qquad \Psi_2=-\lambda_3/2
\label{type1}
\end{equation}
where, in addition, for type D: $\Psi_4=\Psi_0=0$.

For Petrov type II the Segr\'{e} type of $\bf Q$ is [2, 1] with corresponding eigenvalues denoted by $-\lambda_3/2$ and $\lambda_3$ for notational consistency between \eqref{type1} and \eqref{type2}. The frame components of the normal form of $\bf Q$ are
$${\bf Q}= \begin{pmatrix} 1-\lambda_3/2 & -i &0\\-i &1-\lambda_3/2 & 0 \\0 &0& \lambda_3\end{pmatrix} $$ 
Thus, in the associated null tetrad, the NP Weyl tensor components satisfy
\begin{equation}
\Psi_0=\Psi_1=\Psi_3 = 0 \qquad \Psi_4 =-2 \qquad \Psi_2=-\lambda_3/2
\label{type2}
\end{equation}
For the remaining three Petrov types, all the Weyl eigenvalues are zero and so are excluded by the  assumptions (i), (ii) or (iii) of  {\S}1. The corresponding normal forms of $\bf Q$ and $\Psi_i$ are therefore omitted as they will not be needed below.

Thus choosing a suitable null tetrad the  three different sets of assumptions regarding the Weyl eigenvalues in {\S}1 may now be restated as
(where in all cases $\Psi_1=\Psi_3=0$):
\newcounter{M}
\begin{list}
{(\roman{M})}{\usecounter{M}
       \setlength{\rightmargin}{\leftmargin}}
  \item 
  \begin{description} 
  \item[Type I:] $\Psi_2$  and $\Psi_4$ are both non-zero constants;   
  \item[Type D:] $\Psi_2$  is a non-zero constant and $\Psi_0=\Psi_4 = 0$;
  \item[Type II:] $\Psi_2$  is a non-zero constant,  $\Psi_0 = 0$ and $\Psi_4 =-2$.
  \end{description}
  \item $\Psi_2$ is a non-zero constant and $\Psi_4 \neq 0$.
  \item $\Psi_4$ is a non-zero constant multiple of $\Psi_2 (\neq 0)$.
  \end{list}
  
For Petrov type I the choice of a canonical tetrad is not unique; it depends on the choice of numbering of the three spacelike eigenvectors used in \eqref{nul-tet} to construct the complex null tetrad.  As there are six distinct numberings a given Petrov type I field will manifest itself as six different solutions of the Newman-Penrose equations for the spin coefficients and Weyl tensor components  $\Psi_i$. However, these fall naturally into 3 pairs depending on the choice of the spacelike eigenvector used to construct the real null vectors $\ell^a$ and $n^a$. The two members of each pair have the same value of $\Psi_2$ but the values of $\Psi_4$ have opposite signs.

The ambiguity in the numbering of the eigenvectors also means that  type D fields with $\lambda_3=\lambda_1= -\lambda_2/2$  or $\lambda_2=\lambda_3= -\lambda_1/2$ appear as special cases in the analysis of type I spacetimes and need to be excluded. They are characterised by the conditions $\Psi_4 =\pm 3\Psi_2$ respectively.  
Similarly the Petrov type I case where one Weyl eigenvalue is zero appear as special cases in the analysis  and can be immediately excluded by Brans' result\cite{brans}. The cases with $\lambda_1=0$, $\lambda_2=0$ and $\lambda_3=0$ are characterised by the conditions $\Psi_4= \pm \Psi_2$ and $\Psi_2=0$ respectively.

\section{Algebraically Special Spacetimes with Constant Weyl Eigenvalues}
\noindent In this section the type II and type D fields satisfying the assumptions (i) of {\S}1 and {\S}2 are investigated.  It is convenient to consider these two cases together; the NP Weyl tensor components satisfy $\Psi_0=\Psi_1=\Psi_3 = 0$ and $\Psi_2(\neq 0)$ is a constant.
Thus $D\Psi_2$, $\Delta\Psi_2$, $\delta\Psi_2$, $\bar\delta\Psi_2$, $D R$, $\Delta R$, $\delta R$ and $\bar\delta R$  are all zero and for type II $\Psi_4 \ne 0$ whereas for type D, 
$\Psi_4=0$.

In what follows the equation numbers refer to the Newman-Penrose equations in chapter 7  of Stephani et al.\cite{exact-sol}. Using the above restrictions on $\Psi_i$ and their derivatives the Bianchi identities  (7.32a,b,e,h) simplify to 
\begin{equation}
3\kappa\Psi_2=0\qquad 3\sigma\Psi_2=0\qquad 3\rho\Psi_2=0\qquad 3\tau\Psi_2=0
\end{equation}
Hence 
\begin{equation}
\kappa=\sigma=\rho=\tau=0. 
\label{kundt-cond}
\end{equation} 
Thus the spacetime belongs to a special subclass of the Kundt spacetimes\cite{kundt}. Furthermore, the Bianchi identities (7.32f,g) reduce to $3\mu\Psi_2-\sigma\Psi_4=0$ \& $3\pi\Psi_2-\kappa\Psi_4=0$ and so with the aid of \eqref{kundt-cond}:
\begin{equation}
\mu=\pi=0
\label{kcond2}
\end{equation}
These restrictions \eqref{kundt-cond} and \eqref{kcond2} on the spin coefficients depend only on the constancy of $\Psi_2$ and not on the scalings of the null vectors $\ell^a$ and $n^a$ to make $\Psi_4$ constant.

In any scaling of the null tetrad in which $\Psi_4$ is constant the Bianchi identities (7.32c,d) reduce with the aid of \eqref{kundt-cond} to 
\begin{equation}
4\epsilon\Psi_4+3\lambda\Psi_2=0\qquad 4\beta\Psi_4+3\nu\Psi_2=0
\end{equation}
However, these conditions are not used in the analysis  of type II fields below. However for type D where $\Psi_4=0$,  they imply $\nu=\lambda=0$ (as expected by the Goldberg-Sachs theorem\cite{GS} since $n^a$ is also a principal null direction in this case). 
 
Substituting \eqref{kundt-cond} in the Newman-Penrose form of the  Ricci equations (7.21q) one obtains $\Psi_2+R/12=0$ and thus by contracting \eqref{efe}
$\Psi_2=-\Lambda/3)$ or equivalently by \eqref{type1} or \eqref{type2} $\lambda_3=2\Lambda/3$.   Thus under the assumption of constancy, {\it the non-repeated eigenvalue of the Weyl tensor of an algebraically special Einstein spacetime of type D or II necessarily has the value} $2\Lambda/3$.

These Einstein  spaces belong to a special subclass of the Kundt spacetimes\cite{kundt} considered by Lewandowski\cite{lewan}.  In terms of a complex coordinate $z$  and real coordinates $u$ and $v$, the metric may be written in the form 
\begin{equation}
ds^2=2P^{-2}dz d\bar z-2du(dv+Wdz+\bar W d\bar z+Hdu) 
\label{kundt-metric}
\end{equation}
where $P=P(z, \bar z, u)$ and $H=H(z, \bar z, u, v)$ are real and $W=W(z, \bar z, u)$ is complex; $W$ is independent of $v$ 
as a consequence of $\tau=0$. The repeated principal null vector $\ell^a$ is given by $\partial_v$.  The coordinate freedom preserving the form of \eqref{kundt-metric} is given by equations (31.10) of Stephani et al.\/\cite{exact-sol}. As stated by Lewandowski\cite{lewan}, using the coordinate freedom (31.10a) in $z$  the solutions of the field equations for the metric functions $P$, $H$ and $W$ may be written as
\begin{equation}
 P=1+\Lambda z \bar z/2 \qquad H =-\Lambda v^2/2+H_0(z,\bar z, u), \qquad W=iL_{,z}
 \end{equation}
where $L$ is a \emph{real} potential satisfying 
\begin{equation}
P^2 L_{,z\bar z}=-\Lambda L
\label{Leqn}
\end{equation} 
To complete the identification of the algebraically special spacetimes with constant Weyl scalars with those considered by Lewandowski, it is necessary to check that all Lewandowski spacetimes are Petrov type II (or D) with  constant Weyl scalars. Choosing a complex null basis of one-forms:
\begin{subequations}
\label{oneforms}
\begin{eqnarray}
\ell_i dx^i & = & (H+P^2W\bar W)du+dv \\
n_i  dx^i & = & du \\
m_i dx^i & = & -P\bar W du +P^{-1}d z. 
\end{eqnarray}
\end{subequations}
A straightforward calculation using the computer algebra system Classi\cite{Aman} shows that 
\begin{subequations}
\begin{eqnarray}
\Psi_2 &=& \Lambda/3, \qquad \Psi_0=\Psi_1=\Psi_3=0,\\
\Psi_4 &=&-iv\Lambda(P^2L_{,\bar z})_{,\bar z}+i(P^2L_{,\bar z u})_{,\bar z}+(P^2H_{0,\bar z})_{,\bar z}-\Lambda (PL_{,\bar z})^2
\label{Psi4}
\end{eqnarray}
\end{subequations}
so that the null tetrad is a canonical tetrad of the Weyl tensor (apart from a scaling of $\ell^a$ and $n^a$ and a rotation of $m^a$ to make 
$\Psi_4=1$). 
The general solution of equation \eqref{Leqn}  for the potential $L$ is \cite{lewan} 
 \begin{equation}
 L= \Re(\Lambda P^{-1}  \bar zf(z,u) -f_{,z}(z,u))
 \end{equation}
 where $f(z,u)$ is an arbitrary function analytic in $z$.
The remaining field equation implies 
 \begin{equation}
 H_{0,z\bar z}  =\Lambda L_{,z}L_{,\bar z}-\Lambda^2 P^{-2}L^2
 \end{equation}
Given $L$, this can be integrated to give $H_0$ up to addition of an arbitrary harmonic function $\Re h_0(z, u)$.  It can be seen that the general solution of Petrov type II depends on two arbitrary complex functions $f(z,u)$ and $h_0(z, u)$ analytic in $z$. 
 
 For type D the condition $\Psi_4=0$ implies that the the coefficient of $v$ in  \eqref{Psi4}  vanishes. Thus $(P^2L_{,\bar z})_{,\bar z}=0$ and hence  $L_{,\bar z}=g(z, u)/P^2$ for some function $g(z, u)$. This then implies that $f(z, u) = A(u)z^2+B(u)z+C(u)$ where $A, B$ and $C$ are arbitrary functions of $u$.  Using the expression for $L_{,\bar z}$ the term independent of $v$ in \eqref{Psi4}  reduces to 
 $(P^2H_{0,\bar z})_{,\bar z}-\Lambda g(z, u)L_{,\bar z}$. As this also vanishes, it follows that $P^2H_{0,\bar z} = \Lambda g(z, u)L +h(z, u)$  for some function $h(z, u)$. A somewhat messy calculation shows that the remaining coordinate freedom (31.10a, b \& c) preserving the form of the metric can be used to set $H_0 =W=0$.  {\it Thus the type D metric is decomposable into two 2-spaces of constant curvature:}
 \begin{equation}
 ds^2=2(1+\Lambda z\bar z/2)^{-2}dz d\bar z-2dudv-\Lambda v^2du^2.
 \end{equation} 
 and is therefore a Robinson-Bertotti solution\cite{rob, bert}. It is homogenous with a multiply-transitive isometry group of dimension 6.

\section{Algebraically General Spacetimes with Constant Weyl Eigenvalues}
\noindent In this section the type I fields satisfying the assumptions (i) of {\S}1 and {\S}2 are investigated.  The NP Weyl tensor components satisfy $\Psi_1=\Psi_3 = 0$ and  $\Psi_2$ \& $\Psi_4 (=\Psi_0)$  are non-zero constants. Thus $D\Psi_2$, $\Delta\Psi_2$, $\delta\Psi_2$, $\bar\delta\Psi_2$, $D\Psi_4$, $\Delta\Psi_4$, $\delta\Psi_4$ and $\bar\delta\Psi_4$ are all zero. With these restrictions on $\Psi_i$ and their derivatives the Bianchi identities (7.32) of \cite{exact-sol} become purely algebraic equalities:

$$\begin{array}{ll}
(4\alpha-\pi)\Psi_4+3\kappa\Psi_2=0 &  \qquad 3\pi\Psi_2-\kappa\Psi_4 =0  \\ 
(4\gamma-\mu)\Psi_4+3\sigma\Psi_2=0 & \qquad 3\mu\Psi_2-\sigma\Psi_4=0 \\
(4\epsilon-\rho)\Psi_4+3\lambda\Psi_2=0 & \qquad 3\rho\Psi_2-\lambda\Psi_4=0 \\
(4\beta-\tau)\Psi_4+3\nu\Psi_2=0  & \qquad 3\tau\Psi_2-\nu\Psi_4=0
\end{array} $$
These may be solved to express eight of the spin coefficients in terms of the remaining four $\kappa$, $\sigma$, $\nu$ and $\lambda$:
\begin{equation}
\rho = \psi \lambda, \qquad \tau = \psi \nu, \qquad \mu = \psi \sigma, \qquad \pi = \psi \kappa,
\label{spin1}
\end{equation}
\begin{equation}
\alpha = \frac{(\psi^2-1)\kappa}{4\psi}, \qquad \beta = \frac{(\psi^2-1)\nu}{4\psi}, \qquad \epsilon = \frac{(\psi^2-1)\lambda}{4\psi}, 
\qquad \gamma = \frac{(\psi^2-1)\sigma}{4\psi}.
\label{spin2}
\end{equation}
where a subsidiary constant $\psi$  defined by 
\begin{equation}
\label{psi}
\psi = \frac{\Psi_4}{3\Psi_2}\quad {\rm or\ equivalently}\quad \psi = \frac{\Psi_0}{3\Psi_2}\
\end{equation}
has been introduced for later convenience.  The following values of $\psi$ are excluded:
\begin{equation}
\psi \neq 0, \pm 1  {\rm \  (type\ D)} \qquad  \psi \neq  \pm 1/3  {\rm \  (zero\ Weyl\ eigenvalue)}
\end{equation}

Using \eqref{spin1} and \eqref{spin2} these eight spin coefficients may be completely eliminated from the Newman-Penrose  Ricci identities (7.21).  Here and below equation references of the form (7.xx) refer to Chapter 7 of Stephani et al.\cite{exact-sol}.
In particular from (7.21a \& g) two simultaneous equations for $D\lambda$ \& $\bar\delta\kappa$ are obtained which may be solved to yield:
\begin{subequations}
\label{singleDs}
\begin{eqnarray}
D\lambda & = & \frac{-4\psi\bar\psi(\kappa^2+\bar\kappa\nu)+\lambda^2\bar\psi(3+5\psi^2)-\lambda\bar\lambda\psi(1-\bar\psi^2)}
                                      {4\psi\bar\psi} \\
\bar\delta\kappa & = &   \frac{4 \psi\bar\psi(\lambda^2-\sigma\bar\sigma) -\kappa^2\bar\psi(3+5\psi^2) -\kappa\bar\nu\psi(1-\bar\psi^2)}                            
				{4\psi\bar\psi}
\end{eqnarray}
Similarly, from the pairs (7.21c \& i), (7.21k \& m) and (7.21n \& p) the following are obtained:
\begin{eqnarray}
D\nu & = & \frac{-4\psi\bar\psi(\kappa\sigma-\bar\kappa\lambda\bar\psi)+\lambda\nu\bar\psi(3+5\psi^2)+\bar\lambda\nu\psi(1-\bar\psi^2)}
                                      {4\psi\bar\psi} \\
\Delta\kappa & = &   \frac{4 \psi\bar\psi(\lambda\nu-\sigma\bar\nu\bar\psi) -\kappa\sigma\bar\psi(3+5\psi^2) -\kappa\bar\sigma\psi(1-\bar\psi^2)}                            
				{4\psi\bar\psi} \\
\delta\lambda & = & \frac{-4\psi\bar\psi(\kappa\sigma+\nu\bar\lambda\bar\psi)+\lambda\nu\bar\psi(3+5\psi^2)-\bar\kappa\lambda\psi(1-\bar\psi^2)}
                                      {4\psi\bar\psi} \\
\bar\delta\sigma & = &   \frac{4 \psi\bar\psi(\lambda\nu+\kappa\bar\sigma\bar\psi) -\kappa\sigma\bar\psi(3+5\psi^2) +\sigma\bar\nu\psi(1-\bar\psi^2)}                            
				{4\psi\bar\psi} \\
\delta\nu & = & \frac{4 \psi\bar\psi(\lambda\bar\lambda-\sigma^2)+\nu^2\bar\psi(3+5\psi^2)+\bar\kappa\nu\psi(1-\bar\psi^2)}
                                      {4\psi\bar\psi} \\
\Delta\sigma & = &   \frac{4 \psi\bar\psi(\kappa\bar\nu+\nu^2) -\sigma^2\bar\psi(3+5\psi^2) +\sigma\bar\sigma\psi(1-\bar\psi^2)}                            
				{4\psi\bar\psi}
\end{eqnarray}
\end{subequations}
Thus 8 of the 16 derivatives of $\kappa$, $\sigma$, $\nu$ and $\lambda$ are now known.   The NP Ricci identity (7.21e) is identically satisfied as a consequence of \eqref{spin2} and (\ref{singleDs}c, e) as is (7.21r) as a consequence of \eqref{spin2} and (\ref{singleDs}d, f).

The NP Ricci identities (7.21b \& h) are both equalities for $D\sigma-\delta\kappa$ and so eliminating this expression results in a purely algebraic identity:
\begin{equation}
24(\kappa\nu-\sigma\lambda)(1-\psi^2)-12\Psi_2(1-3\psi^2)-R=0.
\label{alg1}
\end{equation}
Similarly (7.21j \& q) are both equalities for $\Delta\lambda-\bar\delta\nu$ and eliminating this expression also results in \eqref{alg1}.
A second algebraic expression may be obtained by subtracting (7.21$\ell$) from (7.21f) multiplying by the result by $4\psi/(\psi^2-1)$ and then adding the difference of (7.21j \& b). The result, after multiplication by $3\psi(\psi^2-1)$, is
\begin{equation}
6(\kappa\nu-\sigma\lambda)(1-\psi^2)(1-5\psi^2)-\psi^2(18\psi^2\Psi_2-42\Psi_2+R)=0.
\label{alg2}
\end{equation}
From \eqref{alg1} and \eqref{alg2} the following simpler algebraic relations may be deduced:
\begin{subequations}
\label{algEqs}
\begin{eqnarray}
\Psi_2 & = & \frac{(\kappa\nu-\sigma\lambda)(9\psi^2-1)}{9\psi^2}\\
R & = & 4(\kappa\nu-\sigma\lambda)(3\psi^2+1)^2/3.
\end{eqnarray}
\end{subequations}
From these two equations it can be immediately deduced that $\kappa\nu-\sigma\lambda \neq 0$ as $\Psi_2 \neq 0$ and that the pure vacuum case ($R =0$) is characterised by $\psi = \pm i/\sqrt{3}$.  Note that for the pure vacuum case the Weyl eigenvalues are proportional to the three cube roots of $-1$; for example $\lambda_1=(-1+i\sqrt{3})\lambda_3/2$ and $\lambda_2=(-1-i\sqrt{3})\lambda_3/2$ for the choice $\psi =+i/\sqrt{3}$.

In addition to the previous two algebraic equations \eqref{algEqs} and the eight equations \eqref{singleDs} for single derivatives of $\kappa, \sigma, \nu$ and $\lambda$, five independent equations involving pairs of derivatives of these spin coefficients remain; for example from (7.21b, d, f, j \& o) on eliminating $\rho, \tau, \mu, \pi, \alpha, \beta, \epsilon, \gamma, R$ and $\Psi_2$ with the aid of  \eqref{spin1}, \eqref{spin2} and \eqref{algEqs}, one obtains:  
\begin{subequations}
\label{doubleDs}
\begin{eqnarray}
D\sigma-\delta\kappa &=&
 \frac{5\bar\psi(3\psi^2+1)(\kappa\nu-\sigma\lambda) +3\psi(3\bar\psi^2+1)(\sigma\bar\lambda+\kappa\bar\kappa)}{12\psi\bar\psi}\\
D\kappa-\bar\delta\lambda&= & 
\frac{2\bar\psi(11\psi^2+1)\kappa\lambda+4\psi\bar\psi(\nu\bar\sigma-\bar\kappa\sigma)+\psi(1-\bar\psi^2)(\bar\nu\lambda-\kappa\bar\lambda)}
 {4\psi\bar\psi}\\
D\sigma-\Delta\lambda&= & 
\frac{8\bar\psi(6\psi^2+1)\kappa\nu+2\bar\psi(9\psi^2-1)\sigma\lambda}{12\psi\bar\psi}  \nonumber\\
& &+\frac{3\psi(1-\bar\psi^2)(\lambda\bar\sigma+\bar\lambda\sigma)+12\psi\bar\psi^2(\kappa\bar\kappa+\nu\bar\nu)}{12\psi\bar\psi}\\
\Delta\lambda-\bar\delta\nu &=& 
-\frac{5\bar\psi(3\psi^2+1)(\kappa\nu-\sigma\lambda)+3\psi(3\bar\psi^2+1)(\lambda\bar\sigma+\nu\bar\nu) }  {12\psi\bar\psi}\\
\delta\sigma-\Delta\nu &=&
\frac{2\bar\psi(11\psi^2+1)\sigma\nu+4\psi\bar\psi(\kappa\bar\lambda-\bar\nu\lambda)+\psi(1-\bar\psi^2)(\bar\kappa\sigma-\nu\bar\sigma)}
 {4\psi\bar\psi}
\end{eqnarray}
\end{subequations}
Note that the equation  for 
$\delta\kappa-\bar\delta\nu$ derived from (7.21$\ell$) is dependent on (\ref{doubleDs}a, c, d).

On using \eqref{spin1} \& \eqref{spin2} the commutator (7.6b) and the complex conjugate of commutator (7.6c) become
\begin{subequations}
\label{Comm}
\begin{eqnarray}
\delta D- D\delta & = & \Bigl(\frac{(\psi^2-1)\nu}{4\psi}-\frac{(3\bar\psi^2+1)\bar\kappa}{4\bar\psi}\Bigr) D+\kappa\Delta\nonumber\\
& &-\Bigl ( \frac{(\psi^2-1)\lambda}{4\psi}+\frac{(3\bar\psi^2+1)\bar\lambda}{4\bar\psi}\Bigr)\delta -\sigma\bar\delta\\
\bar\delta\Delta -\Delta\bar\delta & = & -\nu D + \Bigl(\frac{(3\bar\psi^2+1)\bar\nu}{4\bar\psi}-\frac{(\psi^2-1)\kappa}{4\psi}\Bigr)\Delta\nonumber\\
& &+\lambda\delta +\Bigl(\frac{(3\bar\psi^2+1)\bar\sigma}{4\bar\psi}+\frac{(\psi^2-1)\sigma}{4\psi}\Bigr)\bar\delta
\end{eqnarray}
\end{subequations}
The commutator $\delta D- D\delta $ is now applied to $\lambda$ and various derivatives  are eliminated using  \eqref{singleDs} and \eqref{doubleDs} to produce the following equation involving the two unknown derivatives  of $\lambda$:
\begin{subequations}
\label{DEqs}
\begin{equation}
\kappa\Delta\lambda-\sigma\bar\delta\lambda=(1-\bar\psi^2)\lambda(\sigma\bar\nu-\kappa\bar\sigma)/(4\bar\psi)
+\nu(\sigma\bar\sigma-\bar\psi\kappa\bar\nu)+\kappa(\kappa\nu-\sigma\lambda)/(3\psi).
\end{equation}
Similarly applying the commutator  $\delta D- D\delta $ to $\nu$ yields
\begin{equation}
\kappa\Delta\nu-\sigma\bar\delta\nu=-(1-\bar\psi^2)\nu(\sigma\bar\nu-\kappa\bar\sigma)/(4\bar\psi)
+\lambda(\kappa\bar\nu-\bar\psi\sigma\bar\sigma)+\sigma(\kappa\nu-\sigma\lambda)/(3\psi),
\end{equation}
and the commutator $\bar\delta\Delta -\Delta\bar\delta$ applied to $\sigma$ and $\kappa$ produces
 \begin{equation}
\nu D\sigma-\lambda\delta\sigma = (1-\bar\psi^2)\sigma(\nu\bar\lambda-\bar\kappa\lambda)/(4\bar\psi)
-\kappa(\lambda\bar\lambda-\bar\psi\bar\kappa\nu)-\nu(\kappa\nu-\sigma\lambda)/(3\psi) \vspace{-5 pt}
 \end{equation}
 \begin{equation}
 \nu D\kappa-\lambda\delta\kappa = -(1-\bar\psi^2)\kappa(\nu\bar\lambda-\bar\kappa\lambda)/(4\bar\psi)
-\sigma(\bar\kappa\nu-\bar\psi\lambda\bar\lambda)-\lambda(\kappa\nu-\sigma\lambda)/(3\psi)
\label{Comm4}
\end{equation}
respectively.
\end{subequations}

The equations \eqref{doubleDs} and (\ref{DEqs}a, b, c) may now be solved for the eight remaining derivatives of the spin coefficients:
\begin{subequations}
\label{singleDs2}
\begin{eqnarray}
D\kappa &=& \frac{\kappa\sigma \lambda\bar\psi(19+81\psi^2)-3\kappa\sigma\bar\lambda\psi(1-\bar\psi^2)-16\kappa^2\nu\bar\psi(1+3\psi^2)} {12\sigma\psi\bar\psi}-\bar\kappa\sigma\\
\delta\kappa &=& \frac{\sigma\lambda\bar\psi(1+4\psi^2) -\kappa\nu\bar\psi(3+5\psi^2)-\kappa\bar\kappa\psi(1-\bar\psi^2)} {4\psi\bar\psi}-\sigma\bar\lambda\bar\psi\\
D\sigma &=& \frac{\sigma\lambda\bar\psi(7+33\psi^2) +3\sigma\bar\lambda\psi(1-\bar\psi^2)-4\kappa\nu\bar\psi}{12\psi\bar\psi}+\kappa\bar\kappa\bar\psi\\
\delta\sigma &=&\frac{\sigma\nu\bar\psi(1+11\psi^2)+\bar\kappa\sigma\psi(1-\bar\psi^2)} {4\psi\bar\psi}+\kappa\bar\lambda\\
\Delta\nu &=&\frac{-\sigma\nu\bar\psi(1+11\psi^2)+\bar\sigma\nu\psi(1-\bar\psi^2)} {4\psi\bar\psi}+\bar\nu\lambda\\
\bar\delta\nu &=& \frac{-\kappa\nu\bar\psi(7+33\psi^2)+3\nu\bar\nu\psi(1-\bar\psi^2)+4\sigma\lambda\bar\psi}{12\psi\bar\psi}+\bar\sigma\lambda\bar\psi\\
\Delta\lambda &=& \frac{-4\kappa\nu\bar\psi(1+4\psi^2)+\sigma\lambda\bar\psi(3+5\psi^2)-\bar\sigma\lambda\psi(1-\bar\psi^2)} {4\psi\bar\psi}
-\nu\bar\nu\bar\psi\\
\bar\delta\lambda &=& \frac{\kappa\sigma\lambda\bar\psi(13+15\psi^2)-3\sigma\bar\nu\lambda\psi(1-\bar\psi^2)-16\kappa^2\nu\bar\psi(1+3\psi^2)} {12\sigma\psi\bar\psi} -\bar\sigma\nu
\end{eqnarray}
\end{subequations}
In the above it has  been assumed that $\sigma \ne 0$ since $\sigma$ appears in the denominators of (\ref{singleDs2}a, h); the case 
$\sigma=0$ will be considered below.

Substituting \eqref{singleDs2} in the commutator equation \eqref{Comm4} the following purely algebraic relation is obtained:
\begin{equation}
\label{PureVac}
\bar\psi(1+3\psi^2)(\kappa\nu-\sigma\lambda)^2 =0.
\end{equation}
From this it may be  concluded that $\psi=\pm i/\sqrt{3}$ (since $\psi \ne 0$ as type D is excluded and from the first of \eqref{algEqs} 
$\kappa\nu-\sigma\lambda \ne 0$ since $\Psi_2 \ne 0$). Without loss of generality we may choose $\psi=+i/\sqrt{3}$ as the negative sign simply corresponds to interchanging the  two Weyl eigenvalues $\lambda_1$ and $\lambda_2$.  
  Thus, from (\ref{algEqs}b), {\it the spacetime is a pure vacuum spacetime} ($R=0$ or equivalently $\Lambda=0$). 

\subsection{The case $\sigma=0$}
\noindent In this subsection it is assumed that $\sigma=0$ and hence from the Goldberg-Sachs theorem\cite{GS} $\kappa \ne 0$.  Then from (\ref{singleDs}f) it follows that $\lambda\nu = 0$ and hence $\lambda=0$ (since $\nu=0$ leads to an immediate contradiction since it implies $\Psi_2=0$ from (\ref{algEqs}a)).  Now from (\ref{singleDs}a, h) it follows that 
$\kappa^2+\bar\kappa\nu=0$ and $\nu^2+\kappa\bar\nu=0$. Thus 
\begin{equation}
\label{nu-kappa}
\nu =\epsilon_1 \bar\kappa\qquad{\rm and}\qquad \kappa^2+\epsilon_1 \bar\kappa^2=0
\end{equation}
where $\epsilon_1=\pm 1$. As a consequence of the first of \eqref{nu-kappa}, the equations (\ref{singleDs}c, d) lead to
\begin{equation}
\label{kapDs}
D\kappa=D\nu= \Delta\kappa =\Delta\nu =0
\end{equation}

When $\nu=+\bar\kappa$ substituting  $\sigma=\lambda=0$ in (\ref{doubleDs}c) produces the simple purely algebraic equation $6\psi^2+3\psi\bar\psi +1=0$. This is only consistent if $\psi=\pm i/\sqrt(3)$. On the other hand when $\nu=-\bar\kappa$ simplification of  (\ref{doubleDs}c) produces $6\psi^2-3\psi\bar\psi +1=0$ which is inconsistent.  Thus necessarily $\epsilon_1=1$, $\nu= +\bar\kappa$ and the spacetime is again pure vacuum (as $\psi=\pm i/\sqrt{3}$). 

Without loss of generality we may again choose $\psi=+i/\sqrt{3}$. Then from the second of \eqref{nu-kappa}, 
$\kappa=\pm k(1+\epsilon_2 i)/\sqrt{2}$ where $k=|\kappa|$ and $\epsilon_2 =\pm 1$ is another sign factor.  As a consequence \eqref{singleDs} and \eqref{doubleDs} imply that $\delta\kappa = \bar\delta\kappa=0$.  Thus $\kappa$ is a constant from \eqref{kapDs} and hence {\it all the spin coefficient are constants}.

To summarise when $\sigma =0$ we have a pure vacuum spacetime with
\begin{equation}
\label{sigma0}
\sigma=\lambda=0, \quad \nu =\bar\kappa, \quad \kappa=\pm k(1+\epsilon_2 i)/\sqrt{2},\quad \psi = i/\sqrt{3}, \quad\Psi_2=8k^2/3
\end{equation}
where $k$ is a positive  constant.  Thus all the spin coefficients are constant as a consequence of \eqref{spin1} and \eqref{spin2}; in fact
\begin{equation}
\rho=\mu =\epsilon=\gamma=0,\quad \pi=\alpha = i\kappa/\sqrt{3}, \quad \tau=\beta=i\bar\kappa/\sqrt{3}.
 \end{equation}
 
Together with the result for the case $\sigma\ne 0$  it has been shown that {\it the only Petrov type I Einstein spacetimes with constant Weyl eigenvalues have $\Lambda=0$ (i.e.\ are pure vacuum) and the  Weyl eigenvalues are proportional to  the  three cube roots of $-1$}. Thus, a fortiori, there are no homogeneous proper Einstein spaces of Petrov type I; this recovers a  result of MacCallum \& Siklos\cite{MacSik} without using group theoretic methods.   In the same paper MacCallum \& Siklos showed there are no homogeneous proper Einstein spaces of Petrov type II; a result also confirmed by the analysis in {\S}3 as it is easy to show the Lewandowski metrics are not homogeneous. Note that, unlike the case for Petrov type I, this result for type II does not generalise to the case with constant Weyl eigenvalues.
 
\subsection{The general case $\sigma \ne 0$}
\noindent Again without loss of generality it is assumed below that 
\begin{equation}
\label{psi-vac}
\psi=+i/\sqrt{3}.
\end{equation}
As all 16 derivatives of the spin coefficients $\kappa, \sigma, \nu \ \&\  \lambda$ are now known, the remaining commutators may be applied to these  spin coefficients and all the derivatives eliminated  using equations \eqref{singleDs} and \eqref{singleDs2} to produce a number of purely algebraic compatibility relations.
From the commutator (7.6a) for $\Delta D-D\Delta$ applied to $\sigma, \kappa, \lambda \ \&\ \nu$ the following cubic relations are obtained
\begin{subequations}
\label{cubics}
\begin{eqnarray}
2\kappa\sigma\nu+\sigma^2\lambda+\sigma\bar\kappa\bar\nu-\sigma\bar\sigma\bar\lambda +i\sqrt{3}\nu^2\lambda&=& 0\\ 
2\kappa\sigma\lambda+\kappa^2\nu+\kappa\bar\sigma\bar\lambda-\kappa\bar\kappa\bar\nu+i\sqrt{3}\nu\lambda^2 &=& 0\\
2\kappa\nu\lambda+\sigma\lambda^2+\bar\kappa\bar\nu\lambda-\bar\sigma\lambda\bar\lambda +i\sqrt{3}\kappa^2\sigma &=& 0\\
2\sigma\nu\lambda +\kappa\nu^2+\bar\sigma\nu\bar\lambda-\bar\kappa\nu\bar\nu+i\sqrt{3}\kappa\sigma^2 &=& 0
\end{eqnarray}
respectively. The commutator (7.6d) $\bar\delta\delta-\delta\bar\delta$ applied to $\sigma, \kappa, \lambda \ \&\ \nu$ leads to the same four equations as do the commutators $(\delta\Delta-\Delta\delta)\kappa$, $(\bar\delta D-D \bar\delta)\sigma$, $(\bar\delta D - D\bar\delta)\nu$ and $(\delta\Delta-\Delta\delta)\lambda$. The commutators $(\bar\delta D-D \bar\delta)\kappa$, $(\delta\Delta-\Delta\delta)\nu$,
$(\delta\Delta-\Delta\delta)\sigma$ and $(\bar\delta D - D\bar\delta)\lambda$ yield four more cubic relations:
\begin{eqnarray}
\lambda^3+\sigma\bar\sigma\bar\lambda-\bar\kappa\sigma\bar\nu -i\sqrt{3}\kappa^2\lambda &=& 0\\
\sigma^3  +\bar\sigma\lambda\bar\lambda-\bar\kappa\bar\nu\lambda-i\sqrt{3}\sigma\nu^2 &=& 0\\
 \nu^3+\kappa\bar\kappa\bar\nu-\kappa\bar\sigma\bar\lambda-i\sqrt{3}\sigma^2\nu &=& 0\\
 \kappa^3+\bar\kappa\nu\bar\nu-\bar\sigma\nu\bar\lambda-i\sqrt{3}\kappa\lambda^2 &=& 0
\end{eqnarray}
respectively.
\end{subequations} 
The remaining commutators namely $\delta D - D\delta$ applied to $\kappa$ \& $\sigma$ and $\bar\delta \Delta - \Delta\bar\delta$ applied to $\nu$ \& $\lambda$ are identically satisfied as a consequence of \eqref{singleDs} and \eqref{singleDs2} as, of course, are those in \eqref{DEqs} on substituting for $\psi$ using \eqref{psi-vac}.
\subsubsection{Special cases}
\noindent To solve these cubic equations first consider the special cases where one of the spin coefficients is zero. Firstly the case $\sigma=0$ has already been considered in the previous subsection and, if $\lambda=0$, it follows from (\ref{cubics}e) that $\sigma=0$ (as $\kappa\nu=0$ would imply $\Psi_2=$ from \eqref{algEqs}. Hence it may be assumed below that $\sigma$ and $\lambda$ are non-zero. Next if $\kappa=0$, it follows from (\ref{cubics}h) that $\nu=0$   Similarly if $\nu=0$ it follows from (\ref{cubics}g) that $\kappa=0$.   Thus, when $\kappa=\nu=0$, it follows from (\ref{cubics}a, e) that $\sigma\lambda=\bar\sigma\bar\lambda$ and $\lambda^2 =-\sigma^2$.  Thus $\lambda=\pm i\sigma$ \& 
$\bar\sigma^2=-\sigma^2$. Then from (\ref{singleDs}b) it follows that $\sigma^2+\sigma\bar\sigma=0$ which results in the contradiction
$\sigma^2(1\pm i)=0$.  Thus it may now be assumed below that none of the spin coefficients $\kappa, \sigma, \nu\ \& \ \lambda$ vanish. 

Subtracting $\nu\times(\ref{cubics}b)$ from $\lambda\times(\ref{cubics}a)$ results in 
\begin{equation}
\label{eq1}
\lambda\sigma\bar\kappa\bar\nu-\lambda\sigma\bar\sigma\bar\lambda+\sigma^2\lambda^2+\nu\bar\nu\kappa\bar\kappa-\nu\bar\lambda\kappa\bar\sigma-\nu^2\kappa^2 = 0
\end{equation}
as does subtracting $\kappa\times(\ref{cubics}d)$ from $\sigma\times(\ref{cubics}c)$.
Now $\kappa\times(\ref{cubics}h)-\lambda\times(\ref{cubics}e)-\eqref{eq1}$ and $\nu\times(\ref{cubics}g)-\sigma\times(\ref{cubics}f)-\eqref{eq1}$ produce 
\begin{subequations}
\label{eq34}
\begin{eqnarray}
-\lambda^4+\kappa^4-\sigma^2\lambda^2+\nu^2\kappa^2 &=& 0\\
-\sigma^4+\nu^4-\sigma^2\lambda^2+\nu^2\kappa^2 &= & 0
\end{eqnarray}
\end{subequations}
respectively.  Then adding the two equations \eqref{eq34} yields
\begin{equation}
(\kappa^2+\nu^2-\sigma^2-\lambda^2)(\kappa^2+\nu^2+\sigma^2+\lambda^2) = 0
\end{equation}
Thus 
\begin{equation}
\label{keyeq}
\sigma^2+\lambda^2 =\pm(\kappa^2+\nu^2).
\end{equation}.

If $\sigma^2+\lambda^2 =\kappa^2+\nu^2 =0$, then $\lambda =i\epsilon_1 \sigma$ \& $\nu = i\epsilon_2 \kappa$ where the $\epsilon$'s are independent sign factors with $\epsilon_1^2=\epsilon_2^2=1$. Substituting  for $\nu$ and $\lambda$ in (\ref{cubics}a,  b) and dividing the first by $\sigma$ and the second by $\kappa$ and adding produces the equation 
\begin{equation}
(\epsilon_1+i\sqrt{3}\epsilon_2)\kappa^2+(\epsilon_2+ i\sqrt{3}\epsilon_1)\sigma^2 = 0
\end{equation}
Thus two cases arise. 
\begin{enumerate}
\setlength{\itemsep}{-0.6 ex}
\item If $\epsilon_2=-\epsilon_1, \quad\kappa^2=+\sigma^2\quad$ so that $\quad\lambda =i\epsilon_1\sigma, \quad \kappa=\epsilon_3\sigma, \quad \nu =-i\epsilon_1\epsilon_3\sigma$.
\item If $\epsilon_2=+\epsilon_1, \quad\kappa^2=-\sigma^2\quad$ so that $\quad\lambda =i\epsilon_1\sigma, \quad \kappa=i\epsilon_3\sigma, \quad \nu =-\epsilon_1\epsilon_3\sigma$, 
\end{enumerate}
where $\epsilon_3$ is third sign factor.  In either case $\Psi_2=-8i\epsilon_1\sigma^2/3$ so $\sigma$ is constant.  Then subtracting (\ref{singleDs}a \& h) and simplifying yields in both cases $2i\sigma\bar\sigma/\sqrt{3}=0$ and so $\sigma=0$ which is a contradiction.
\subsubsection{Generic case}
\noindent Now equation \eqref{keyeq} may be written as  $\sigma^2+\lambda^2 =\epsilon_1(\kappa^2+\nu^2) \ne 0$ where $\epsilon_1=\pm 1$.  Now dividing each of  \eqref{eq34} by $\sigma^2+\lambda^2$ and using \eqref{keyeq} produces the following pair of equations: $\lambda^2=\epsilon_1\kappa^2$ and 
$\nu^2=\epsilon_1\sigma^2$. Substitution of these relations in \eqref{eq1} and factorising yields 
$$(\sigma\lambda+\kappa\nu)(\bar\sigma\bar\lambda-\bar\kappa\bar\nu)=0$$
The second factor cannot vanish (otherwise $\Psi_2=0$) and so $\sigma\lambda=-\kappa\nu)$. As a consequence two case arise.
\begin{subequations}
\label{cases}\
\begin{eqnarray}
\lambda=\epsilon_2\kappa,& \qquad \nu =-\epsilon_2\sigma\qquad &  {\rm when\ }\epsilon_1=+1 \\
\lambda=i\epsilon_2\kappa,& \qquad \nu =-i\epsilon_2\sigma\qquad &  {\rm when\ }\epsilon_1=-1
\end{eqnarray}
\end{subequations}
where $\epsilon_2=\pm 1$ is a second sign factor. 

Substituting for $\lambda$ and $\nu$ in (\ref{cubics}a) and dividing by $\sigma$ the following equations are obtained in cases (a) and (b) respectively:
\begin{subequations}
\label{kap-sig}
\begin{eqnarray}
\bar\kappa\bar\sigma &=& \frac{-1+i\sqrt{3}}{2}\kappa\sigma \\
\bar\kappa\bar\sigma &=& \frac{1+i\sqrt{3}}{2}\kappa\sigma 
\end{eqnarray}
\end{subequations}
Now substituting  for $\lambda$ and $\nu$ in (\ref{cubics}f), using \eqref{kap-sig} to eliminate $\bar\kappa\bar\sigma$ and then dividing by $\sigma$ the following equations are obtained in the two cases respectively: 
\begin{subequations}
\label{sig}
\begin{eqnarray}
(\kappa^2-\sigma^2)(1-i\sqrt{3}) = 0 &\quad {\rm so\ that} & \quad \sigma=\epsilon_3\kappa \\
(\kappa^2+\sigma^2)(1+i\sqrt{3}) = 0 &\quad {\rm so\ that} & \quad \sigma=i\epsilon_3\kappa
\end{eqnarray}
\end{subequations}
where $\epsilon_3 =\pm 1$ is another sign factor.
Now eliminating  $\lambda$, $\nu$ and $\sigma$ in (\ref{cubics}c) and (\ref{algEqs}a) yields the following equations:
\begin{subequations}
\label{kappa}
\begin{eqnarray}
\bar\kappa^2 = \frac{-1+i\sqrt{3}}{2}\kappa^2\quad & &\Psi_2 = -\frac{8\epsilon_2\epsilon_3\kappa^2}{3} \\
\bar\kappa^2 = -\frac{1+i\sqrt{3}}{2}\kappa^2 \quad & & \Psi_2 = +\frac{8\epsilon_2\epsilon_3\kappa^2}{3} 
\end{eqnarray}
\end{subequations}
 The solutions of the preceding equations \eqref{kappa}  for $\kappa$ are
\begin{subequations}
\label{kappa2}
\begin{eqnarray}
\kappa =& \pm k(-\sqrt{3}+i)/2 \quad &{\rm or} \quad \kappa =\pm k(1+i\sqrt{3})/2\\
\kappa =& \pm k(\sqrt{3}+i)/2 \quad &{\rm or} \quad \kappa =\pm k(1-i\sqrt{3})/2 
\end{eqnarray}
\end{subequations}
where $k=|\kappa|$. It also follows from the equalities for $\Psi_2$ in \eqref{kappa} that $\kappa$ {\it  is constant and hence so is $k$ and all the other spin coefficients}.

If the first alternative in (\ref{kappa2}a) namely $\kappa = \pm k(-\sqrt{3}+i)/2$ is substituted in (\ref{singleDs}a) and then (\ref{cases}a) \& (\ref{sig}a) used to eliminate $\lambda, \nu$ \& $\sigma$, the equation $k^2(\epsilon_2\epsilon_3-1)=0$ is obtained which is clearly only consistent when $\epsilon_3= \epsilon_2$. With this relation it is found that all the remaining equations in \eqref{singleDs} \& \eqref{singleDs2} are satisfied.  However, if the second alternative for $\kappa$  in (\ref{kappa2}a) is used  in the simplification process of (\ref{singleDs}a),  the equation $k^2(\epsilon_2\epsilon_3+1 -2i/\sqrt{3})=0$ is obtained which is clearly inconsistent. 

Repeating the same process with  (\ref{cases}b) \& (\ref{sig}b) and the first alternative  in (\ref{kappa2}b), this time leads to an immediate inconsistency in  (\ref{singleDs}a).  However, repeating the simplification process with the second alternative  in (\ref{kappa2}b) namely $\kappa = \pm k(1-i\sqrt{3})/2$  it transpires that \eqref{singleDs} \& \eqref{singleDs2} are again only satisfied if  $\epsilon_3= \epsilon_2$.

 To summarise there are essentially three pure vacuum cases in all of which the spin coefficients are constant.
 \begin{subequations}
 \label{solns}
 \begin{eqnarray}
  \sigma=\lambda=0, &\nu =+\bar\kappa, &  \kappa=\pm k(1+\epsilon_2 i)/\sqrt{2}, \quad\Psi_2=8k^2/3\\
  \lambda=\sigma=\epsilon_2\kappa, & \nu =-\kappa, & \kappa = \pm k(-\sqrt{3}+i)/2, \quad \Psi_2=4k^2(-1+i\sqrt{3})/3\\
 \lambda=\sigma=i\epsilon_2\kappa, & \nu =+\kappa, & \kappa = \pm k(1-i\sqrt{3})/2, \quad \Psi_2=-4k^2(1+i\sqrt{3})/3
 \end{eqnarray}
 \end{subequations}
where $k$ is an arbitrary positive constant and $\epsilon_2=\pm 1$. In all three cases the remaining spin coefficients are given by 
\begin{eqnarray}
\rho = i \lambda/\sqrt{3}, \qquad \tau = i\nu/\sqrt{3},& \quad \mu = i \sigma/\sqrt{3},& \quad \pi = i\kappa/\sqrt{3}.\\
\alpha = i\kappa/\sqrt{3}, \qquad \beta =i\nu/\sqrt{3},& \quad \epsilon =i\lambda/\sqrt{3}, &\quad \gamma = i\sigma/\sqrt{3}.
\end{eqnarray}
As discussed in {\S}2 the choice of a canonical tetrad is not unique; it depends on which of the ordering of the  spacelike eigenvectors  used in \eqref{nul-tet} to construct the complex null tetrad.  Thus we expect there to be three pairs of distinct solutions of the Newman-Penrose equations for the spin coefficients and Weyl tensor components  $\Psi_i$ corresponding to the same spacetime. The choice of $\psi=+i/\sqrt{3}$ effectively singles out one member of each pair and hence the appearance of three solutions in \eqref{solns} is to be expected.  The fourfold sign ambiguity in each of these solutions is also to be expected as the vectors of the orthonormal eigentetrad  $(e_1^a, e_2^a, e_3^a, u^a)$ are only determined up to sign. Some of these sign changes correspond to interchange of $\ell^a$ and $n^a$ and/or $m^a$ and $\bar m^a$. A simultaneous sign change of $u^a $ and $e_3^a$ changes the sign of both $\ell^a$ and $n^a$ which reverses the signs of $\sigma$ and $\lambda$, but leaves $\kappa$ and $\nu$ unaltered. On the other hand a  simultaneous sign change of $e_1^a $ and $e_2^a$ changes the sign of $m^a$  which reverses the signs of $\kappa$ and $\nu$, but leaves $\sigma$ and $\lambda$ unaltered.

Thus it is expected that the three cases in \eqref{solns} all correspond to the same underlying spacetime.  In fact,  the spacetime in question is homogeneous with metric:
\begin{equation}
\label{hom1}
 ds^2= \frac{1}{4k^2}\biggl(dx^2+e^{-4x/\sqrt{3}}dy^2+e^{2x/\sqrt{3}}\cos(2x)(dz^2-dt^2)-2e^{2x/\sqrt{3}}\sin(2x)dz dt\biggr)  
 \end{equation} 
This metric is originally due to Petrov\cite{petrov2} who derived it using group theoretic methods. The coordinates have been scaled compared to those in {\S}12.2 of Stephani et al. \cite{exact-sol} for consistency with the notation used in this paper. 
A  canonical orthonormal tetrad of Weyl eigen-one-forms corresponding to  case (a) in \eqref{solns}  is:
\begin{subequations}
\begin{eqnarray}
u_adx^a &= & \frac{e^{x/\sqrt{3}} }{2k}\bigl( \cos x\ dt - \sin x\ dz \bigr)\\
e_{3a}dx^a &= & \frac{e^{x/\sqrt{3}}} {2k}\bigl( \sin x\ dt +\cos x\ dz \bigr)\\
e_{1a}dx^a &=&  \frac {1}{ 2\sqrt{2}k}\bigl( dx -e^{-2x/\sqrt{3}}dy \bigr)\\
e_{2a}dx^a &=& \frac {1}{ 2\sqrt{2}k}\bigl( dx +e^{-2x/\sqrt{3}} dy \bigr).
\end{eqnarray}
\end{subequations}
and the associated complex null tetrad is given by \eqref{nul-tet}.  A straightforward calculation using the computer algebra system Classi\cite{Aman} shows that 
$$\Psi_0=\Psi_3=0,\quad \Psi_0=\Psi_4 = 8ik^2/\sqrt{3}, \quad \Psi_2=8k^2/3$$ 
The metric \eqref{hom1} is the only homogeneous Einstein space of Petrov type I \cite{petrov2}.  The analysis in this paper establishes the stronger result that it is {\it the only Einstein space of Petrov type I with constant Weyl eigenvalues}.

 \section{One Constant Weyl Eigenvalue}
 \noindent In this section we consider  Petrov type I Einstein spaces with only one constant Weyl eigenvalue; without loss of generality this is taken to be $\lambda_3$. As in {\S}4 
 $\Psi_1=\Psi_3=0$ and $\Psi_0=\Psi_4$ and $\Psi_2$ is a non-zero constant.  It is convenient to introduce the quantity $\psi$ as in equation\eqref{psi} but, of course, $\psi$ is not constant in this case.  The values $0, \pm 1, \pm 1/3$ of $\psi$ are excluded as usual. With these assumptions the NP Bianchi identities (7.32c, d, e \& f) of \cite{exact-sol} become purely algebraic identities:
 \begin{equation}
 \rho=\psi\lambda,\quad\tau=\psi\nu,\quad\pi=\psi\kappa,\quad\mu=\psi\sigma
  \label{spin4}
  \end{equation}
 Thus the spin coefficients $\rho, \tau, \mu$ \& $\pi$ may be eliminated from the Ricci identities and remaining Bianchi identities.
 The Bianchi identities (7.32g, b, h \& a) become, on dividing by $3\Psi_2$ and simplifying:
 \begin{subequations}
 \label{BIs2} 
 \begin{eqnarray}
 D\psi &=&(\psi^2-1)\lambda-4\psi\epsilon \\
 \Delta\psi &=&-(\psi^2-1)\sigma+4\psi\gamma \\
 \delta\psi &=&(\psi^2-1)\nu-4\psi\beta \\
 \bar\delta\psi &=&-(\psi^2-1)\kappa+4\psi\alpha
 \end{eqnarray}
 \end{subequations}
 respectively.
 
Substituting \eqref{spin4} in  the Ricci identities (7.21) of \cite{exact-sol}, the combination $(h)-\psi\times(b)$ reduces to the purely algebraic equation:
\begin{equation}
\kappa\nu-\sigma\lambda=\frac{(1-3\psi^2)\Psi_2+R/12}{2(1-\psi^2)}
\label{cross-prod2}
\end{equation}
which is equivalent to \eqref{alg1} of {\S}4.   The combination $(q)-\psi\times(j)$ of Ricci identities simplifies to the same result. Using \eqref{psi} and $R = 4\Lambda$ this equation may be written as
\begin{equation}
\kappa\nu-\sigma\lambda=3\frac{\Psi_2(3-\Psi_0\Psi_4/\Psi_2^2)+\Lambda}{2(9-\Psi_0\Psi_4/\Psi_2^2)}
\end{equation}
This equation is invariant under boosts and rotations of the canonical null tetrad:
\begin{equation}
\tilde\ell^a=A\ell^a,\qquad\tilde n^a=A^{-1}n^a,\qquad\tilde m^a =e^{i\theta}m^a.
\label{boost-rot}
\end{equation}

\section{Petrov Type I with $\Psi_4$ a Constant Multiple of $\Psi_2$}
\noindent  In this section Petrov type I spacetimes where the Weyl  eigenvalues are constant multiples of a single (complex) scalar field are
 considered. This condition is equivalent to the assumption that,  in a null eigen-tetrad of the Weyl tensor, $\Psi_4$ is a  constant 
 multiple of $\Psi_2$.   As in {\S}4 $\Psi_1=\Psi_3=0$ and $\Psi_0=\Psi_4$ and it is convenient to introduce the constant $\psi$ as in equation\eqref{psi}.  The values $0, \pm 1$ of $\psi$ are excluded as the Weyl 
 tensor is Petrov type D in these cases and similarly the values $\pm 1/3$ are ruled out as they correspond to the case of a zero Weyl eigenvalue which is excluded by Brans' theorem\cite{brans}.

Eliminating $\Psi_0$ \& $\Psi_4$  using \eqref{psi} the Bianchi identities (7.32 c, d, e, f) of \cite{exact-sol} simplify to
\begin{subequations} 
\label{BIs}
\begin{eqnarray}
D \Psi_2 &=& 3(\rho-\psi\lambda)\Psi_2\\
 \delta\Psi_2 &=&  3(\tau-\psi\nu)\Psi_2\\
\bar \delta\Psi_2 &= & 3(\psi\kappa-\pi)\Psi_2\\
\Delta \Psi_2 &=& 3(\psi\sigma-\mu)\Psi_2
\end{eqnarray}
\end{subequations}
The Bianchi identities (7.32 g, h, a \& b)  respectively  lead to alternative equations for these four derivatives of $\Psi_2$.  Eliminating the derivative of $\Psi_2$ from corresponding pairs of equations, four purely algebraic equalities are obtained:
\begin{subequations} 
\label{spin3}
\begin{eqnarray}
\alpha &=& \frac{3\psi^2-1}{4\psi}\kappa-\frac{\pi}{2}\\
\beta &=& \frac{3\psi^2-1}{4\psi}\nu-\frac{\tau}{2}\\
\gamma &=& \frac{3\psi^2-1}{4\psi}\sigma-\frac{\mu}{2}\\
\epsilon &=& \frac{3\psi^2-1}{4\psi}\lambda- \frac{\rho}{2}
\end{eqnarray}
\end{subequations}
Thus the four spin coefficients  $\alpha, \beta, \gamma\ \&\ \epsilon$ and their derivatives may be eliminated from the NP Ricci identities (7.21).

The combination $(q)-(h)+2\times (\ell) - 2\times(f)$ of the Ricci identities (7.21) results in an equation in which the only terms in which derivatives occur are 
$$(\Delta\lambda-\bar\delta\nu + \delta\kappa-D\sigma)(3\psi\bar\psi-1)/(2\bar\psi)$$
These combinations of derivatives also appear in (7.21j \& b) and so may be eliminated to produce the purely algebraic equation:
\begin{equation}
\label{Psi2}
\Psi_2 = (\kappa\nu-\sigma\lambda)\frac{9\psi^2 - 1}{9\psi^2}
\end{equation}
which is identical to (\ref{algEqs}a) although here $\Psi_2$ and $\kappa\nu-\sigma\lambda$ are  not constants.  Using \eqref{psi} and rearranging this equation may be rewritten as
\begin{equation}
\kappa\nu-\sigma\lambda = \frac{ \Psi_2}{1-\Psi_2^2/(\Psi_0\Psi_4)}.
\end{equation}
This equation is invariant under boosts and rotations of the canonical null tetrad \eqref{boost-rot}.
 
\section{Conclusions}
Vacuum spacetimes possibly with a non-zero cosmological constant $\Lambda$ (that is Einstein spacetimes) with  constant  non-zero Weyl eigenvalues have been considered. For type II \& D this assumption is sufficient to allow one to prove that the non-repeated eigenvalue necessarily has the value $2\Lambda/3$ and so algebraically special pure vacuum spacetimes of this type are ruled out.  It is then shown that the only possible spacetimes are some Kundt-waves considered by Lewandowski which are type II and  a Robinson-Bertotti solution of type D.  The latter is the direct product of two 2-dimensional spaces of constant curvature and is homogeneous admitting a multiply transitve isometry group of dimension six.  The Lewandowski solutions depend on an arbitrary complex function $f(z, u)$ analytic in $z(=x+iy)$ \& u and an arbitrary real function of $x, y$ \& $u$ harmonic in $x$ and $y$.  These solutions are not homogeneous and in general, they admit no isometries.   Originally these solutions were found by Lewandowski \cite{lewan} by considering Kundt solutions with a reduced holonomy group; this paper therefore gives a new characterisation of these solutions namely the constancy of their Weyl eigenvalues.  It is interesting that the assumption of constancy of the Weyl eigenvalues for Petrov type II \& D implies that  the non-repeated eigenvalue necessarily has the value $2\Lambda/3$.  Many years ago Yakupov \cite{yak1} showed that all  Einstein spaces of embedding class two with non-zero torsion (if such there be) must have a Weyl eigenvalue with this value. Any algebraically special counterexamples to Yakupov's claim \cite{yak2} that Einstein spacetimes of embedding class two with torsion do not exist, will necessarily belong to the Lewandowski class.

For Petrov type I  the only solutions in which all three Weyl eigenvalues are constant must have $\Lambda=0$; there are no proper Einstein spaces satisfying this assumption. Thus {\it a fortiori} there are no homogenous proper Einstein spaces of Petrov type I.  This provides an independent  proof of the result of MacCallum \& Siklos \cite{MacSik} (see also \cite{exact-sol} {\S}12.9) which does not use group theoretic methods.   The only metric turns out to be the homogeneous pure vacuum solution found long ago by Petrov \cite{petrov2} using group theoretic methods. It is the only homogenous  Einstein spacetime of Petrov type I; this paper shows that it can be characterised uniquely by the weaker assumptions of the constancy of the Weyl eigenvalues.

The above results can be summarised in an alternative way by the statement that the only vacuum spacetimes with constant Weyl eigenvalues are either homogeneous or are Kundt spacetimes (of the Lewandowski class). This result is similar to that of Coley et al.\/\cite{coley} who proved their  result for {\bf general} spacetimes under the assumption that all scalar invariants constructed from the curvature tensor {\bf and all its derivatives} were constant.  The result in this paper is restricted to Einstein spaces of Petrov types I, II \& D only, but subject to the weaker assumption of the constancy of the scalar invariants constructed from the curvature tensor alone.

It is somewhat disappointing that despite the seeming generality of the initial assumptions (namely the constancy of the Weyl eigenvalues) that all solutions in this class had previously been found by other methods.  However, the paper does provide new and different characterisations of these solutions based on assumptions which are seemingly weaker than those originally used to derive them.  In particular the method of proof makes no assumptions regarding the isometry or holonomy groups admitted by the spacetimes.

In the final two sections of the paper Petrov type I Einstein spacetimes are considered in which the assumption of the constancy of all three Weyl eigenvalues is weakened in two different ways.  In {\S}5 only one Weyl eigenvalue is assumed constant whilst in {\S}6 the Weyl eigenvalues are assumed to be constant multiples of a single complex scalar field.  These sections are `work in progress'.  However in both cases there is a simple algebraic relation relating what might be called the cross-ratio of the spin coefficients $\kappa, \nu \sigma$ \& $\lambda$,  namely$\kappa\nu-\sigma\lambda$ and a quantity constructed from the curvature.  In {\S}5 the relation is 
\begin{equation}
\kappa\nu-\sigma\lambda=3\frac{\Psi_2(3-\Psi_0\Psi_4/\Psi_2^2)+\Lambda}{2(9-\Psi_0\Psi_4/\Psi_2^2)}
\end{equation}
whereas in {\S}6 the cross-ratio is given by
\begin{equation}
\kappa\nu-\sigma\lambda = \frac{ \Psi_2}{1-\Psi_2^2/(\Psi_0\Psi_4)}
\end{equation}
These equations are invariant under boosts and rotations of the canonical null tetrad \eqref{boost-rot} and so the quantities on each side of the equation are characteristic of the curvature and its canonical null tetrad (rather than artefacts of a particular tetrad scaling). However, their full physical significance is as yet unclear. 

As one might expect both systems of equations are much richer than the case in {\S}4 and are apparently not overdetermined.  These two systems of equations and their integrability conditions are currently being investigated further to see if  any solutions can be found either in the general case or if further assumptions are made. In particular the pure vacuum case ($\Lambda=0$) and the case $\Psi_2=-\Lambda/3$ of relevance to the embedding class two problem will be considered.  Other special cases also seem tractable; for example under the assumptions of {\S}5 the case where the null vector $\ell^a$ of the Weyl canonical null tetrad has zero shear ($\sigma=0$) it is easy to show that the null vector $n^a$ also has zero shear ($\lambda=0$).

\section*{Acknowledgements}
\noindent The extensive calculations in sections 4, 5 and 6 were performed using  Maple package for the Newman-Penrose formalism which was kindly supplied to me by its author Norbert van den Bergh of the University of Ghent.  I would also like to thank him for useful discussions  on this problem and its relation to Yakupov's results for embedding class two spacetimes.

\end{document}